\documentclass[universe,article,accept,pdftex,moreauthors]{Definitions/mdpi} 
\def\hl{}

\firstpage{1} 
\pubvolume{11}
\issuenum{11}
\articlenumber{365}
\pubyear{2025}
\copyrightyear{2025}
\externaleditor{László Árpád Gergely} % More than 1 editor, please add `` and '' before the last editor name
\datereceived{30 September 2025} 
\daterevised{3 November 2025} % Comment out if no revised date
\dateaccepted{4 November 2025} 
\datepublished{6 November 2025} 
\hreflink{https://doi.org/10.3390/\linebreak~universe11110365} % If needed use \linebreak
\doinum{10.3390/universe11110365}
\usepackage{bm}
\def\vx{{\bm x}}
\def\vg{{\bm g}}
\def\vp{{\bm p}}
\def\vL{{\bm L}}
\Title{Hidden Momentum and the Absence of the Gravitational Spin Hall Effect in a Uniform Field}

\TitleCitation{Hidden Momentum and the Absence of the Gravitational Spin Hall Effect in a Uniform Field}

\Author{{Andrzej} %MDPI: Please carefully check the accuracy of names and affiliations. %MDPI: We added an asterisk after this author's name to mark the corresponding author. Please confirm.
 Czarnecki *\orcidA{} and Ting {Gao *} %MDPI: Please check and confirm whether this is the corresponding author, since this information is not consistent with that submitted online at susy.mdpi.com.
 \orcidB{}}

\AuthorNames{Andrzej Czarnecki and Ting Gao}

\AuthorCitation{Czarnecki, A.; Gao, T.}

\address[1]{%
\hl{Department} %MDPI: We merged the affiliations and removed the affiliation numbers since there is only one address. Please confirm.
 of Physics, University of Alberta, Edmonton, AB T6G 2E1, Canada}

\corres{\hangafter=1 \hangindent=1.05em \hspace{-0.82em}{Correspondence: andrzejc@ualberta.ca (A.C.); tgao7@ualberta.ca (T.G.)}}

\abstract{We re-examine the recent claim that a Dirac particle freely falling in a uniform gravitational field exhibits a spin-dependent transverse deflection (gravitational spin Hall effect). Using a circulating mass model, we show that hidden momentum arises in uniform fields when an object carries angular momentum. On the quantum side, we analyze the Dirac Hamiltonian in a uniform potential, construct its Foldy--Wouthuysen form, and evaluate the Heisenberg evolution of spin-polarized Gaussian packets.
The state used previously, with $\langle p\rangle =0$, is not at rest: because canonical and kinetic momenta differ, the packet carries a spin-dependent hidden momentum from $t=0$. Imposing $\langle x(0)\rangle =\langle v(0)\rangle=0$ requires a compensating spin-dependent $\langle p(0)\rangle$; with this preparation $\langle x(t)\rangle =0$ to leading order in the gravitational acceleration $g$. Generalizing, an exact Foldy--Wouthuysen transformation (linear in $g$ but to all orders in $1/c$) shows that spin-dependent transverse motion begins no earlier than at $O(g^2)$ for a broad class of wave packets. We conclude that a uniform field does not produce a gravitational spin Hall effect at linear order; the previously reported drift stems from inconsistent initial states and misinterpreting canonical momentum.}


\begin{document}

%%%%%%%%%%%%%%%%%%%%%%%%%%%%%%%%%%%%%%%%%%
%\setcounter{section}{-1} %% Remove this when starting to work on the template.

\section{Introduction}
Spin--orbit phenomena offer a probe of how internal angular momentum
couples to background fields, and gravity is no exception
\cite{Mathisson:1937zz,Mathisson:2010ui,Papapetrou:1951pa,Dixon:1970zza,Obukhov:2000ih,Silenko:2004ad}:
curvature (encoded via the spin connection in the curved-space Dirac
equation) can imprint helicity-dependent transverse shifts---often
called the gravitational spin Hall effect (SHE)---with analogues in
optics~\cite{onoda2004hall,hosten2008obs}, whose counterpart,
``spinoptics'' in curved spacetime, is still under active debate~\cite{Frolov:2024ebe, Frolov:2024qow, Andersson:2023bvw,Deriglazov:2021gwa}. A
gravitational SHE would test the universality of free fall for
spin degrees of freedom and semiclassical wave-packet dynamics
\cite{Xiao:2009rm, Silenko:2004ad,Obukhov:2000ih} and would have
implications for precision interferometry near Earth and for
polarization-dependent lensing in astrophysics
\cite{noh2021quantum,Andersson:2023bvw}. The uniform-field limit is
the critical baseline because most laboratory settings operate there;
any transverse spin drift in this limit would be both experimentally
accessible and conceptually consequential. Recent claims of a
linear-in-time, polarization-dependent deflection for Dirac wave
packets in a uniform field therefore warrant scrutiny.

\begin{figure}[H]
%\centering
\includegraphics[scale=0.15]{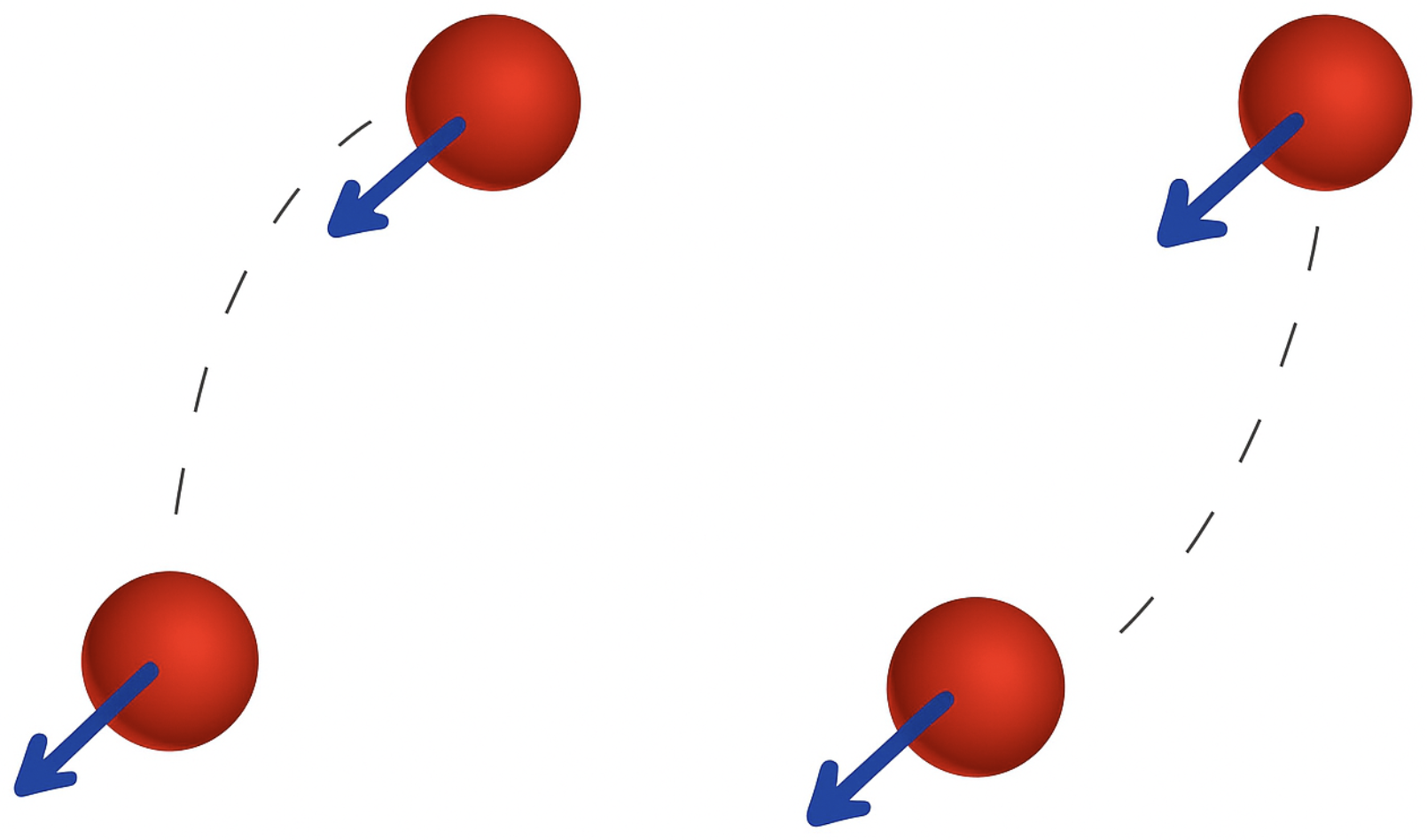}\\[1mm]
(\textbf{a}) \hspace*{45mm} (\textbf{b})
\caption{Two different types of trajectories for the motion of a particle with spin in a uniform gravitational field. (\textbf{a}) Free-fall motion with an initial horizontal velocity; (\textbf{b}) motion starting from rest, with the horizontal velocity developed through the gravitational spin Hall effect. Based on Equations~(22)--(24) in Ref.~\cite{Wang:2023bmd}, the particle should follow a projectile trajectory as in (\textbf{a}); however, the graphic in that work shows trajectory (\textbf{b}), contradicting those equations.
\label{fig:effectWang}}
\end{figure}

A tantalizing recent paper~\cite{Wang:2023bmd} claims that a Dirac
particle freely falling in a constant, uniform gravitational field
(such as near the Earth's surface) may be deflected in a horizontal
direction. Specifically, as shown in Figure~\ref{fig:effectWang},
a particle polarized along a horizontal $y$-axis and falling in the
positive $z$-direction is deflected toward the $\pm x$ direction, depending
on the sign of its polarization.

Here, we expand on a comment we submitted for consideration of publication~\cite{comment:2025} and discuss phenomena related to hidden momentum. In Section~\ref{Wang}, we summarize the reasoning presented in~\cite{Wang:2023bmd}. In Section~\ref{sec:Hidden}, we present a simple model of circulating masses in a uniform gravitational field. We focus on various contributions to the linear momentum. Having developed this intuition, we return to the motion of a spinning particle in Section~\ref{LO}, and in Section~\ref{sec:abs} we demonstrate the absence of the gravitational spin Hall effect in a uniform field.

\section{Summary of Wang's Study of the Gravitational Spin Hall Effect}
\label{Wang}

The analysis presented in Ref.~\cite{Wang:2023bmd} considers a state that, at the initial time
$t=0$, is described by a Gaussian wave \hl{packet,} %MDPI: Please recheck all the equations and make sure there are no duplicated equations in the whole manuscript. %MDPI: Please carefully check variable formatting (italic, bold, subscript, uppercase, etc.) throughout the manuscript to ensure the formatting is consistent and revise if needed..

\begin{equation}
\psi\left(\vx,t=0\right)=\frac{1}{\left(a^{2}\pi\right)^{3/4}}\exp\left(-\frac{\vx^{2}}{2a^{2}}\right),\label{eq:Wang13}
\end{equation}
where $a$ is a constant characterizing the spatial extension of the
initial packet. The complete wave function also contains a Dirac spinor
factor describing polarization along the $\pm y$-axis; it remains
constant during free fall, so we omit it here. 

The time evolution of the wave packet is governed by the Dirac Hamiltonian
with a linear potential term $V\left(z\right)=-mgz$ describing the
interaction with the gravitational field. This Hamiltonian is unitarily
transformed using the Foldy--Wouthuysen \mbox{transformation \cite{Foldy:1949wa,FoldyOriginFW}}.
It is found that at time $t>0$ the expectation values of the $x$
and $z$ coordinates are nonzero. For the $z$ coordinate,
this is, of course, as expected for free fall,
\begin{equation}
\left\langle z\left(t\right)\right\rangle _{\pm}=\frac{gt^{2}}{2},
\end{equation}
where the subscript on the LHS refers to the $y$-polarization state.
However, the $x$ coordinate is also found to develop a nonzero,
polarization-dependent expectation value:
\begin{equation}
\left\langle x\left(t\right)\right\rangle _{\pm}=\mp\frac{\hbar gt}{4mc^{2}}.\label{eq:Wang22}
\end{equation}
\hl{We} %MDPI: Please check if this kind of indentation can be retained since the paragraph starts with a capital letter and revise if necessary. Same below.
 use SI units rather than setting $\hbar=c=1$ to emphasize the
order of quantum and relativistic effects. 

The nonzero expectation value of a horizontal deflection is interpreted
as a gravitational spin Hall effect, previously established only in
non-uniform fields, such as those experienced by a particle flying by a
star~\cite{Liberman:1992zz,Frolov:2024ebe,Frolov:2024qow,Andersson:2023bvw,yamamoto2018spin,bliokh2015quantum,noh2021quantum}.
An astonishing feature of Equation~\eqref{eq:Wang22} is
that the expectation value grows linearly with time. In other words,
the $x$-component of velocity seems to be independent of time; it is nonzero
even at $t=0$ when the wave packet is released. This contradicts
the graphic in Ref.~\cite{Wang:2023bmd}. A motion with a
constant horizontal velocity corresponds to a projectile trajectory depicted
in Figure~\ref{fig:effectWang}a. However, for a phenomenon to
be interpreted as a Hall effect, the particle should develop the transverse
component of velocity from zero. The canonical momentum operator has a zero expectation value in the
initial wave packet described by Equation~\eqref{eq:Wang13}. However, canonical and kinetic momenta differ in the presence of a gravitational field. As we demonstrate below, that packet does have a nonvanishing velocity.

\section{Hidden Momentum}
\label{sec:Hidden}

In this section, we revisit a classical toy model that demonstrates that an object with angular momentum can have ``hidden momentum'' in a gravitational field, which does not vanish even when there is no overall motion of the system.

\subsection{A Simple Model}

The notion of hidden momentum first appeared in a 1967 paper
by Shockley and James~\cite{Shockley67} in the context of electromagnetism.
A concise discussion and a list of key references up to 2012 can be
found in~\cite{griffiths2012resource}.

\begin{figure}[H]
%\centering\includegraphics[scale=0.3]{frameModel.png}
%\centering
\includegraphics[scale=0.6]{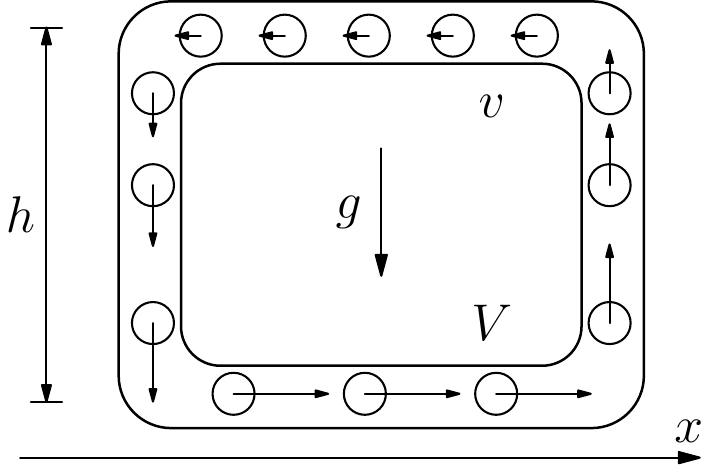}

\caption{\hl{Masses} %MDPI: 1. We moved this figure after its first citation. Please confirm. 2. Please confirm whether it is necessary to add explanations for the English letters and balls in Figure 2. If necessary, please provide explanations. If not necessary, please reply.
 circulating in a frictionless pipe bent into a rectangle. The
rectangle is placed in the constant gravitational field near the Earth.\label{fig:Circul}}
\end{figure}

A simple model of a non-electromagnetic system containing hidden momentum
can be constructed, following an electric model described in~\cite{babson2009hidden}.
Consider a set of masses $m$ that circulate without friction in a pipe
bent into a rectangular shape, shown in Figure~\ref{fig:Circul}. In
the upper horizontal section, $n$ masses progress from right to left
with speed $v$; they accelerate in the left vertical section due
to the gravitational field; $N$ of them move with speed $V>v$ in
the lower horizontal section (conservation of the current imposes
$NV=nv$); and they finally slow down in the right vertical section. In
the nonrelativistic limit, the total mechanical momentum of the circulating
masses vanishes because the larger speed in the lower section is compensated
for by the proportionally smaller number of masses. The net momentum of
the masses arises due to the difference in relativistic factors $\gamma$
corresponding to $v$ and $V$; we denote this effect by a subscript
sr for special relativity,
\begin{align}
\Delta_{\text{sr}}p_{x} & =N\gamma_{V}mV-n\gamma_{v}mv\label{eq:px}\\
 & =N\left(\gamma_{V}mV-\frac{V}{v}\gamma_{v}mv\right)\\
 & \simeq NmV\frac{V^{2}-v^{2}}{2c^{2}}.
\end{align}
\hl{If} the length of the vertical sides is $h$, we have
\begin{equation}
V^{2}-v^{2}=2gh,
\end{equation}
and
\begin{equation}
\Delta_{\text{sr}}p_{x}=NmV\frac{gh}{c^{2}}.\label{eq:pxSR}
\end{equation}

\subsection{Effect of Gravitational Time Dilation}

\textls[-40]{In Equation~\eqref{eq:px}, we used just the special-relativistic factors
$\gamma$, for example,} \mbox{$\gamma_{v}=1/\sqrt{1-v^{2}/c^{2}}$}. Accounting
for the gravitational redshift corrects them by an altitude-dependent
amount,
\begin{equation}
\frac{1}{\sqrt{1-v^{2}/c^{2}}}\to\frac{1}{\sqrt{\left(1+\frac{gz}{c^{2}}\right)^{2}-v^{2}/c^{2}}}\simeq1-\frac{gz}{c^{2}}+\frac{v^{2}}{2c^{2}}.
\end{equation}
\hl{Taking} the center of the circuit as the reference point $z=0$, such
that the upper horizontal section is at $z=h/2$, we get a redshift
correction to the $x$-momentum, 
\begin{equation}
\Delta_{\text{rs}}p_{x}=N\left(\frac{gh}{2c^{2}}mV+\frac{V}{v}\frac{gh}{2c^{2}}mv\right)=N\frac{gh}{c^{2}}mV. 
\label{eq:rs}
\end{equation}
\hl{This} equals the $p_{x}$ found in Equation~\eqref{eq:pxSR}, so it doubles
that effect.

\subsection{Angular Momentum in the Pipe}

The masses in the horizontal sections carry angular momentum relative
to the center of the frame, pointing out of the figure:
\begin{equation}
L_{h}=m\frac{h}{2}\left(V+\frac{V}{v}v\right)N=mhVN.
\end{equation}
\hl{The} masses in vertical sections have speed $u=\left(v+V\right)/2$ on
average. Their number is $N\cdot V/u\cdot h/b$, where $b$ is the
length of a horizontal section, so they contribute to the angular~momentum
\begin{equation}
L_{v}=2\cdot\frac{b}{2}\cdot mu\frac{NVh}{ub}=mNVh=L_{h}.
\end{equation}
\hl{Thus,} the total mechanical \hl{momentum} %MDPI: Please check if the bold formatting in some of the equations and variables is necessary. If not, please remove it.
 $\vp$, $\left|\vp\right|=\Delta_{\text{sr}}p_{x}+\Delta_{\text{rs}}p_{x}$,
can be written in terms of the angular momentum as
\begin{equation}
\vp_{\text{mech}}=\frac{\vL\times\vg}{c^{2}},\label{eq:pmech}
\end{equation}
where $\left|\vL\right|=L_{v}+L_{h}=2mNVh$ and points out of the figure.

\section{Gravitational SHE in the Uniform Gravitational Field Revisited}
\label{LO} 

In Section~\ref{Wang}, we see that the motion of the Dirac particle in the uniform gravitational field derived in~\cite{Wang:2023bmd} leads to unphysical consequences. To understand the origin of this problem, we evaluate the evolution of various operators for the wave packet considered in~\cite{Wang:2023bmd}. Focusing on the upper-left block of the Hamiltonian which describes positive energy states, the FW-transformed Hamiltonian can be written as 
\begin{equation}
    \mathcal{H}_{\mathrm{FW}}=Vmc^2+\frac{\boldsymbol{p}\cdot \left(V\boldsymbol{p}\right)}{2m}+\frac{g\hbar}{4mc^2}\left(\sigma_1p_2-\sigma_2p_1\right),\quad V=1-\frac{gz}{c^2}.
    \label{Hamiltonian, LO}
\end{equation}
\hl{The} wave function considered in~\cite{Wang:2023bmd} is a Gaussian wave packet with the spinor polarized along the positive or negative $y$-direction
\begin{equation}
    \psi_\pm(\boldsymbol{x},0)=\phi(\boldsymbol{x},0)\chi_\pm,
    \label{Gaussian wave packet with spin}
\end{equation}
\begin{equation}
    \phi(\boldsymbol{x},0)=\frac{1}{\left(a^2\pi\right)^{3/4}}e^{-\frac{\boldsymbol{x}^2}{2a^2}},
    \label{Gaussian wave packet}
\end{equation}
\begin{equation}
    \chi_+=\frac{1}{\sqrt{2}}\begin{pmatrix}
1\\
i
\end{pmatrix},\quad 
\chi_-=\frac{1}{\sqrt{2}}\begin{pmatrix}
i\\
1
\end{pmatrix}.
\end{equation}
\hl{The} expectation value of an operator $\mathcal{O}$ at a given time is then
\begin{equation}
\langle\mathcal{O}(t)\rangle=\langle\psi(\boldsymbol{x},0)|U^{-1}\mathcal{O}U|\psi(\boldsymbol{x},0)\rangle,
\end{equation}
where $U=\mathrm{Exp}\left(-i\mathcal{H}_{\mathrm{FW}}t/\hbar\right)$ is the time evolution operator. Using the Baker--Campbell--Hausdorff (BCH) expansion, $U^{-1}\mathcal{O}U$ can be written as
\begin{equation}
U^{-1}\mathcal{O}U=\mathcal{O}+i\frac{t}{\hbar}\left[\mathcal{H}_{\mathrm{FW}},\mathcal{O}\right]-\frac{t^2}{2\hbar^2}\left[\mathcal{H}_{\mathrm{FW}},\left[\mathcal{H}_{\mathrm{FW}},\mathcal{O}\right]\right]+\cdots.
\label{Baker-Hausdorff lemma}
\end{equation}
\hl{By} evaluating $\langle x\rangle$, the author of~\cite{Wang:2023bmd} concluded that the wave packet is moving along the $x$-direction in a spin-dependent way. However, if one evaluates the momentum,
\begin{equation}
    \left[\mathcal{H}_{\mathrm{FW}},p_1\right]=0, 
\label{eq:comzero}
\end{equation}
it leads to $p_1(t)=0$ and suggests the wave packet is not moving along the $x$-direction, in direct conflict with Equation~\eqref{eq:Wang22}. This apparent contradiction is resolved by noting that the momentum in the Hamiltonian refers to the canonical momentum, which does not always reflect the true motion of the particle. Indeed, applying the classical Hamilton's equations to the Hamiltonian in Equation~\eqref{Hamiltonian, LO} gives
\begin{equation}
    m\frac{dx}{dt}=m\frac{\partial \mathcal{H}_{\mathrm{FW}}}{\partial p_1}=Vp_1-\frac{g\hbar}{4c^2}\sigma_2.
\label{eq:23}
\end{equation}
\hl{Therefore,} the conditions for $v_1=0$ and $p_1=0$ are different for the Hamiltonian under consideration. For a particle to initially stay at rest, it has to have a nonzero, spin-dependent initial momentum. This classical observation carries over to quantum mechanics by noting~that
\begin{equation}
    \frac{d}{dt}\langle x(t)\rangle=\langle\psi(\boldsymbol{x},t)|\frac{i}{\hbar}\left[\mathcal{H}_{\mathrm{FW}},x\right]|\psi(\boldsymbol{x},t)\rangle=\langle\psi(\boldsymbol{x},t)| \left(\frac{Vp_1}{m}-\frac{g\hbar}{4mc^2}\sigma_2\right)|\psi(\boldsymbol{x},t)\rangle.
\label{eq:24}
\end{equation}
\hl{Contrary} to the usual expectation, this suggests that the Gaussian wave packet in \mbox{Equation \eqref{Gaussian wave packet},} which is usually believed to be motionless, is moving in the simultaneous presence of spin and a gravitational field. Moreover, under the condition $\langle \boldsymbol{p}(0)\rangle=0$, different orientations of the spin give different initial velocities. To determine whether there is a spin Hall effect, it is important to start from the correct initial conditions. Given that the spin Hall effect refers to the deflection of the trajectory in the presence of spin, it is appropriate to choose $\langle \boldsymbol{x}(0)\rangle=0$ and $\langle \boldsymbol{v}(0)\rangle=0$ as the starting point to discuss the trajectory. For this reason, instead of using the wave functions in Equation~\eqref{Gaussian wave packet with spin}, we consider the wave functions
\begin{equation}
    \Psi_\pm(\boldsymbol{x},0)=\Phi_\pm(\boldsymbol{x},0)\chi_\pm,
\label{eq:25}
\end{equation}
\begin{equation}
    \Phi_\pm(\boldsymbol{x},0)=\frac{1}{\left(a^2\pi\right)^{3/4}}e^{-\frac{\boldsymbol{x}^2}{2a^2}+ ik_\pm x/\hbar},
\end{equation}
where $k_\pm=\pm\frac{g\hbar}{4c^2}$, which introduces a spin-dependent initial momentum to enforce the initial condition $\langle \boldsymbol{v}(0)\rangle=0$. Indeed, as one can explicitly check,
\begin{equation}
    \langle\boldsymbol{v}(0)\rangle_+=\langle\boldsymbol{v}(0)\rangle_-=0.
\label{eq:27}
\end{equation}
\hl{The} trajectory of the particle is then determined by evaluating $\langle\boldsymbol{x}(t)\rangle$ with the help of Equation~\eqref{Baker-Hausdorff lemma}. The motion along the $y$-direction trivially vanishes at leading order in $g$, and the motion along the $z$-direction gives the usual free-fall motion. For the $x$-direction, we get
\begin{equation}
    \left[\mathcal{H}_{\mathrm{FW}},x\right]=\left(\frac{Vp_1}{m}-\frac{g\hbar}{4mc^2}\sigma_2\right),
\label{eq:28}
\end{equation}
whose expectation value for the wave function in Equation~\eqref{eq:25} vanishes exactly as required by the initial conditions. Taking another commutator with $\mathcal{H}_{\mathrm{FW}}$ introduces an extra power of $g$; therefore, we conclude that there is no gravitational spin Hall effect for the motion described by the Hamiltonian in Equation~\eqref{Hamiltonian, LO}.

Before concluding this section, we note that there is some ambiguity in the interpretation of $x$ as the position operator of the particle; however, as discussed at the end of Section~\ref{sec:abs}, this conclusion is not affected.

\section{Absence of Gravitational SHE in a Uniform Gravitational Field: All Orders in $1/c$} \label{sec:abs} 
In this section, we generalize the result in Section~\ref{LO} to linear order in $g$ but to all orders in $1/c$ and for a wide class of wave packets. For this purpose, we start from the Dirac~Hamiltonian
\begin{equation}
    \mathcal{H}=\beta mc^2V+V\boldsymbol{\alpha}\cdot \boldsymbol{p}c-\frac{i\hbar c}{2}\boldsymbol{\alpha}\cdot\boldsymbol{\nabla}V,
    \label{Dirac Hamiltonian}
\end{equation}
where $\beta=\gamma^0$ and $\alpha^i=\gamma^0\gamma^i$. As noted in~\cite{Obukhov:2000ih}, this Hamiltonian allows for an exact FW transformation~\cite{eriksen1960canonical,nikitin1998exact}. Following the discussion in~\cite{eriksen1960canonical}, this is achieved by a two-step unitary transformation
\begin{equation}
    U=U_2 U_1,\quad U_1=\frac{1}{\sqrt{2}}\left(1+\gamma^0 J\right), \quad U_2=\frac{1}{\sqrt2}\left(1+\gamma_0\epsilon'\right),
\end{equation}
where $J$ is a unitary operator that anticommutes with both $\mathcal{H}$ and $\gamma^0$:
\begin{equation}
    JJ^\dagger=J^\dagger J=1, \quad J\mathcal{H}=-\mathcal{H}J,\quad J\gamma^0=-\gamma^0 J.
    \label{J requirements}
\end{equation}
\hl{A} Hamiltonian in general contains both odd (i.e., anticommuting with $\gamma^0$) and even pieces (i.e., commuting with $\gamma^0$):
\begin{equation}
    \mathcal{H}=\mathcal{H}_{\mathrm{odd}}+\mathcal{H}_{\mathrm{even}},\quad \mathcal{H}_{\mathrm{odd}}=\frac{1}{2}\left(\mathcal{H}-\gamma^0\mathcal{H}\gamma^0\right),\quad \mathcal{H}_{\mathrm{even}}=\frac{1}{2}\left(\mathcal{H}+\gamma^0\mathcal{H}\gamma^0\right).
\end{equation}
\hl{The} effect of $U_1$ is to transform this Hamiltonian to a new one that contains only odd terms,
\begin{equation}
    \mathcal{H}'=U_1\mathcal{H}U_1^\dagger=\mathcal{H}_{\mathrm{odd}}-J\gamma^0\mathcal{H}_{\mathrm{even}}, \quad \gamma^0 \mathcal{H}'=-\mathcal{H}'\gamma^0.
\end{equation}
\hl{The} second transformation is then defined in terms of this new Hamiltonian:
\begin{equation}
    \mathcal{H}_{\mathrm{FW}}=U_2\mathcal{H}'U_2^\dagger,\quad \epsilon'=\mathcal{H}'/\sqrt{(H')^2},
\end{equation}
and this transformation converts the odd Hamiltonian into an even one:
\begin{equation}
    \mathcal{H}_{\mathrm{FW}}=\gamma^0\sqrt{(\mathcal{H}')^2}.
\label{eq:35}
\end{equation}
\hl{The} key challenge for constructing the exact FW transformation is to find the operator $J$ that satisfies the requirements in Equation~\eqref{J requirements}. For Hamiltonian \eqref{Dirac Hamiltonian} under consideration, this is possible with
\begin{equation}
    J=i\gamma^0\gamma^5,
\end{equation}
where $\gamma^5=i\gamma^0\gamma^1\gamma^2\gamma^3$. We therefore find
\begin{equation}
    \mathcal{H}'=V\gamma^0\boldsymbol{\gamma}\cdot\boldsymbol{p}c-imc^2V\gamma^0\gamma^5-\frac{i\hbar c}{2}\gamma^0\boldsymbol{\gamma}\cdot\boldsymbol{\nabla}V,
\end{equation}
\begin{adjustwidth}{-\extralength}{0cm}
\begin{equation}
    \mathcal{H}_{\mathrm{FW}}=\gamma^0\sqrt{m^2c^4V^2+Vp^2c^2V+\frac{\hbar^2 c^2}{2}V\Delta V-\frac{\hbar^2c^2}{4}\phi^2+V\hbar c^2\boldsymbol{\Sigma}\cdot\left(\boldsymbol{\phi}\times\boldsymbol{p}\right)+m\hbar c^3V\boldsymbol{\phi}\cdot\boldsymbol{\gamma}\gamma^5},
\end{equation}
\end{adjustwidth}
where
\begin{equation}
    \boldsymbol{\phi}=\boldsymbol{\nabla}V,\quad \boldsymbol{\Sigma}=
\begin{pmatrix}
\boldsymbol{\sigma} & 0\\
0 & \boldsymbol{\sigma}
\end{pmatrix}.
\end{equation}
\hl{Focusing} on the upper-left block of the Hamiltonian and taking the uniform gravitational potential $V=1-gz/c^2$, this Hamiltonian simplifies:
\begin{equation} \mathcal{H}_{\mathrm{FW}}=\sqrt{m^2c^4V^2+Vp^2c^2V-gV\hbar \boldsymbol{\sigma}\cdot\left(\hat{\boldsymbol{z}}\times\boldsymbol{p}+mc\hat{\boldsymbol{z}}\right)-\frac{\hbar^2}{4c^2}g^2}.
\label{eq:40}
\end{equation}
\hl{To} understand the behavior of this Hamiltonian, we take the dominant term $m^2c^4$ outside the square root and expand in the relativistic and gravitational corrections:
\begin{align}
    \mathcal{H}_{\mathrm{FW}}&=mc^2\sqrt{1+\xi},
    \\
    \xi &=V^2-1+\frac{1}{m^2c^2}Vp^2V-\frac{1}{m^2c^4}gV\hbar\boldsymbol{\sigma}\cdot\left(\hat{\boldsymbol{z}}\times\boldsymbol{p}+mc\hat{\boldsymbol{z}}\right)-\left(\frac{\hbar g}{2mc^3}\right)^2.
\end{align}
\hl{Expanding} the Hamiltonian for small $\xi$ generates powers $\xi^n$, which in general are difficult to calculate. However, if one focuses on terms up to linear order in $g$ (under the assumptions $gz/c^2\ll 1$ and $g\hbar/(mc^3)\ll 1$), it is possible to proceed relatively easily. In that case, $\xi$ reduces to
\begin{equation}
    \xi=\frac{1}{m^2c^2}Vp^2V-\frac{2gz}{c^2}-\frac{1}{m^2c^4}g\hbar\boldsymbol{\sigma}\cdot\left(\hat{\boldsymbol{z}}\times\boldsymbol{p}+mc\hat{\boldsymbol{z}}\right),
\end{equation}
and after dropping all terms of higher order \hl{in} %MDPI: Please confirm if all the equation labels can be centered.
 $g$,
\begin{align}
    \xi^n=&\frac{1}{(mc)^{2n}}\left(Vp^2V\right)^{n}-\frac{1}{(mc)^{2(n-1)}}\sum_{l=0}^{n-1} p^{2l}\frac{2gz}{c^2}p^{2(n-1-l)}
    \nonumber \\
    &-\frac{ng\hbar}{(mc)^{2n}c^2}p^{2(n-1)}\boldsymbol{\sigma}\cdot\left(\hat{\boldsymbol{z}}\times\boldsymbol{p}+mc\hat{\boldsymbol{z}}\right)
\end{align}
\hl{Using} the relation
\begin{equation}
    p^{2l}zp^{2m}=zp^{2(l+m)}-2i\hbar lp_3p^{2(l+m-1)},
\end{equation}
the result above simplifies to
\begin{adjustwidth}{-\extralength}{0cm}
\begin{equation}
\begin{aligned}
    \xi^n=&\frac{p^{2n}}{(mc)^{2n}}-\frac{2ngz}{c^2}\left[\frac{p^{2n}}{(mc)^{2n}}+\frac{p^{2(n-1)}}{(mc)^{2(n-1)}}\right]\\
    &+\frac{2ni\hbar g p_3}{m^2c^4}\left[n\frac{p^{2(n-1)}}{(mc)^{2(n-1)}}+(n-1)\frac{p^{2(n-2)}}{(mc)^{2(n-2)}}\right]-\frac{ng\hbar\boldsymbol{\sigma}\cdot\left(\hat{\boldsymbol{z}}\times\boldsymbol{p}+mc\hat{\boldsymbol{z}}\right)}{m^2c^4}\frac{p^{2(n-1)}}{(mc)^{2(n-1)}}.
\end{aligned}
\end{equation}
\end{adjustwidth}
\hl{Then,} the Hamiltonian can be written as
\begin{align}
\mathcal{H}_{\mathrm{FW}}= & mc^2\sum_{n=0}^{\infty}a_n\frac{p^{2n}}{(mc)^{2n}}+b_n\frac{gz}{c^2}\frac{p^{2n}}{(mc)^{2n}}+c_n\frac{i\hbar g p_3}{m^2c^4}\frac{p^{2n}}{(mc)^{2n}}
\nonumber \\
&+d_n\frac{g\hbar\boldsymbol{\sigma}\cdot\left(\hat{\boldsymbol{z}}\times\boldsymbol{p}+mc\hat{\boldsymbol{z}}\right)}{m^2c^4}\frac{p^{2n}}{(mc)^{2n}},
    \label{Hamiltonian, all orders}
\end{align}
where
\begin{equation}
\begin{aligned}
    &a_n=\frac{(-1)^{n-1}(2n)!}{4^n(n!)^2(2n-1)},\quad b_n=-2na_n-2(n+1)a_{n+1},\\
    &c_n=2(n+1)^2a_{n+1}+2(n+1)(n+2)a_{n+2},\quad d_n=-(n+1)a_{n+1}.
\end{aligned}
\end{equation}

We now show that the Hamiltonian in Equation~\eqref{Hamiltonian, all orders} does not lead to the spin Hall effect up to linear order in $g$. For this purpose, we evaluate $\langle x(t)\rangle$ under the initial conditions $\langle x(0)\rangle=0$ and $\frac{d}{dt}\langle x(t)\rangle|_{t=0}=\langle \psi(\boldsymbol{x},0)|i\left[\mathcal{H}_{\mathrm{FW}},x\right]|\psi(\boldsymbol{x},0)\rangle/\hbar=0$. The commutator~$\left[\mathcal{H}_{\mathrm{FW}},x\right]$~gives
\begin{equation}
\begin{aligned}
    \left[\mathcal{H}_{\mathrm{FW}},x\right]=mc^2\sum_{n=1}^{\infty}&\left(-2ni\hbar p_1\right)\left(a_n\frac{p^{2(n-1)}}{(mc)^{2n}}+b_n\frac{gz}{c^2}\frac{p^{2(n-1)}}{(mc)^{2n}}+c_n\frac{i\hbar g p_3}{m^2c^4}\frac{p^{2(n-1)}}{(mc)^{2n}}\right.\\
    &\left.+d_n\frac{g\hbar\boldsymbol{\sigma}\cdot\left(\hat{\boldsymbol{z}}\times\boldsymbol{p}+mc\hat{\boldsymbol{z}}\right)}{m^2c^4}\frac{p^{2(n-1)}}{(mc)^{2n}}\right)-mc^2\sum_{n=0}^{\infty}d_n\frac{ig\hbar^2\sigma_y}{m^2c^4}\frac{p^{2n}}{(mc)^{2n}}.
\end{aligned}
\label{eq:48}
\end{equation}

When taking more commutators with $\mathcal{H}_{\mathrm{FW}}$, we note that the spin-dependent, non-commuting terms involve either $g\left[p_3,gz\right]$ or $g^2\left[\sigma_i,\sigma_j\right]$; they are all of order $O(g^2)$ or higher. Using Equation~\eqref{Baker-Hausdorff lemma}, we get
\begin{equation}
   \langle x(t)\rangle_\mathrm{spin-dependent}\sim O(g^2),
\label{eq:49}
\end{equation}
as long as the wave function does not introduce extra $1/g$ behavior. A similar argument leads to $\langle y(t)\rangle_\mathrm{spin-dependent}\sim O(g^2)$, and $\left[\mathcal{H}_{\mathrm{FW}},z\right]$ does not contain the spin. Therefore, all spin-dependent motion necessarily starts at order $O(g^2)$ or higher. We note that this result is consistent with the semiclassical analysis of the vector model of spin~\cite{Ramirez:2017pmp}.

In the above analysis, we show that there is no gravitational spin Hall effect in a uniform gravitational field up to linear order in $g$ for the Hamiltonian in Equation~\eqref{Hamiltonian, all orders}. Before closing the discussion, we comment on the applicability of this conclusion in other representations. If one takes the first few terms in Equation~\eqref{Hamiltonian, all orders}, the result is
\begin{equation}
    \mathcal{H}_{\mathrm{FW,LO}}=mc^2\left(1-\frac{gz}{c^2}\right)+\frac{p^2}{2m}-\frac{g\hbar \boldsymbol{\sigma}\cdot\left(\hat{\boldsymbol{z}}\times\boldsymbol{p}+mc\hat{\boldsymbol{z}}\right)}{2mc^2}.
\label{eq:50}
\end{equation}
\hl{Comparing} with Equation~\eqref{Hamiltonian, LO}, the form of the spin-dependent interaction is different. As noted in~\cite{Obukhov:2000ih}, this Hamiltonian is related to the one in Equation~\eqref{Hamiltonian, LO} by a unitary transformation, which changes not only the form of the Hamiltonian but also the position and spin operators. Indeed, it is a nontrivial task to interpret the meaning of these operators in a specific representation~\cite{Foldy:1949wa,Newton:1949cq,fradkin1961electron,Silenko:2004ad}. A more recent discussion of this issue as well as a potential framework for the analysis can be found in~\cite{Deriglazov:2017jub}. However, our argument for the vanishing of the gravitational spin Hall effect to leading order in $g$ essentially relies only on three pieces of information:
\begin{enumerate}
    \item The expectation value of commutator $\left[\mathcal{H}_{\mathrm{FW}},\boldsymbol{x}\right]$ vanishes as a result of the requirement that the particle start from rest.
    \item In the $g\to 0$ limit, the Hamiltonian and the wave function reduce to those for free~particles.
    \item The spin-dependent interaction is independent of the position of the particle, which is true for the uniform gravitational field.
\end{enumerate}

Items 2 and 3 together guarantee that the spin-dependent part in $\left[\mathcal{H}_{\mathrm{FW}},\left[\mathcal{H}_{\mathrm{FW}},\boldsymbol{x}\right] \right]$ (and higher-order commutators) starts at order $g^2$. Therefore, although we consider the Hamiltonian in~\cite{Obukhov:2000ih}, the same conclusion applies to a much wider class of representations, for example, the Hamiltonian in~\cite{Silenko:2004ad}, as well.

\section{Summary}\label{sec:summary}
We revisited the proposed gravitational spin Hall effect in a
\emph{uniform} field and tracked both classical and quantum sources of
apparent transverse motion. Classically, the circulating mass model shows that the
hidden relativistic mechanical momentum
\mbox{(Equations \eqref{eq:px}--\eqref{eq:pxSR} and \eqref{eq:rs})} can be written as
\( \vp_{\mathrm{mech}}=\vL\times\vg/c^{2}\)
(Equation~\eqref{eq:pmech}). Quantum mechanically, starting from the FW
Hamiltonian in a linear potential, Equation~\eqref{Hamiltonian, LO}, we noted that the
canonical transverse momentum is conserved,
\([H_{\mathrm{FW}},p_{1}]=0\) \mbox{(Equation~\eqref{eq:comzero}),} while the velocity operator
contains a spin term, \(m\,\dot{x}=Vp_{1}-\tfrac{g\hbar}{4c^2}\sigma_{2}\)
\mbox{(Equation~\eqref{eq:23})}, implying \(d\langle x\rangle/dt=\langle
Vp_{1}/m-\tfrac{g\hbar}{4mc^2}\sigma_{2}\rangle\) (Equation~\eqref{eq:24}). Therefore, a
packet with \(\langle p(0)\rangle=0\) is \emph{\hl{not} %MDPI: Please confirm if the italics are necessary; if not, please remove them.
} at rest. Preparing
truly at-rest states requires a spin-dependent phase with \(k_{\pm}=\pm
g\hbar/4c^2\) in the initial wave functions (Equations~\eqref{eq:25}--\eqref{eq:27}), which enforces
\(\langle[H_{\mathrm{FW}},x]\rangle=0\) at \(t=0\) (Equation~\eqref{eq:28}) and
eliminates any \(O(g)\) transverse drift. Generalizing, we used the
exact FW transformation (Equations~\eqref{Dirac Hamiltonian}--\eqref{eq:35}) and its specialization to
\(V=1-gz/c^2\) \mbox{(Equation \eqref{eq:40})} to derive the all-orders-in-\(1/c\),
linear-in-\(g\) Hamiltonian (Equation~\eqref{Hamiltonian, all orders}). The commutator structure then
shows that spin-dependent transverse motion does not arise at orders
lower than \( g^{2}\) (Equations~\eqref{eq:48} and \eqref{eq:49}). While the leading-order Hamiltonian
\(H_{\mathrm{FW,LO}}\) (Equation~\eqref{eq:50}) is related to
Equation~\eqref{Hamiltonian, LO} by a unitary transformation,
the absence of a uniform-field gravitational SHE at \(O(g)\) is
representation-independent. Our result shows consistency with the expectation from the equivalence principle. The previously reported drift \(\langle
x(t)\rangle_{\pm}=\mp \hbar gt/(4mc^{2})\) (Equation~\eqref{eq:Wang22}) is traced to
inconsistent initial conditions and conflicting canonical and kinetic
momenta.

%%%%%%%%%%%%%%%%%%%%%%%%%%%%%%%%%%%%%%%%%%
\vspace{6pt} 
%%%%%%%%%%%%%%%%%%%%%%%%%%%%%%%%%%%%%%%%%%
\authorcontributions{\hl{Conceptualization, A.C. and T.G.; Methodology, A.C. and T.G.; Formal analysis, A.C. and T.G.; Investigation, A.C. and T.G.; Writing---original draft, A.C. and T.G.; Writing---review and editing, A.C. and T.G.} %MDPI: We have revised the Author Contributions statement according to the system, please confirm.
 \hl{All authors} %MDPI: We moved this mandatory statement into a separate sentence. Please confirm.
 have read and agreed to the published version of the~manuscript.}

\funding{\hl{This research was funded by the Natural Sciences and Engineering Research Council of Canada (NSERC).}} %MDPI: Please provide the grant numbers.

\dataavailability{\hl{No new data were created or analyzed in this study. Data sharing is not applicable to this article.}} %MDPI: We noticed that the Data Availability Statement mentioned in the manuscript is different from the system. Please confirm.

\acknowledgments{During the preparation of this study, the authors used ChatGPT 5 for the purposes of generating graphics. The authors have reviewed and edited the output and take full responsibility for the content of this publication.}

\conflictsofinterest{The authors declare no conflicts of interest. The funders had no role in the design of the study; in the collection, analyses, or interpretation of data; in the writing of the manuscript; or in the decision to publish the results.}

\begin{adjustwidth}{-\extralength}{0cm}
%\centering %% If there is a figure in wide page, please release command \centering, for Table, ``\textwidth" should be ``\fulllength"

\reftitle{References}

\PublishersNote{}
\end{adjustwidth}

\begin{thebibliography}{999}
\bibitem[Mathisson(1937)]{Mathisson:1937zz}
Mathisson, M.
\newblock {Neue mechanik materieller systemes}.
\newblock {\em Acta Phys. Polon.} {\bf 1937}, {\em 6},~163--200.
\bibitem[Mathisson(2010)]{Mathisson:2010ui}
Mathisson, M.
\newblock Republication of: New mechanics of material systems.
\newblock {\em Gen. Relativ. Gravit.} {\bf 2010}, {\em
42},~1011--1048. [\href{http://doi.org/10.1007/s10714-010-0939-y}{CrossRef}]
\bibitem[Papapetrou(1951)]{Papapetrou:1951pa}
Papapetrou, A.
\newblock {Spinning test particles in general relativity. \hl{I.} 
}
\newblock {\em Proc. R. Soc. Lond. A} {\bf 1951}, {\em 209},~248--258. [\href{http://dx.doi.org/10.1098/rspa.1951.0200}{CrossRef}]
\bibitem[Dixon(1970)]{Dixon:1970zza}
Dixon, W.G.
\newblock {Dynamics of extended bodies in general relativity. I. Momentum and
angular momentum}.
\newblock {\em Proc. R. Soc. Lond. A} {\bf 1970}, {\em 314},~499--527. [\href{http://dx.doi.org/10.1098/rspa.1970.0020}{CrossRef}]
\bibitem[Obukhov(2001)]{Obukhov:2000ih}
Obukhov, Y.N.
\newblock {Spin, gravity, and inertia}.
\newblock {\em Phys. Rev. Lett.} {\bf 2001}, {\em 86},~192--195. [\href{http://dx.doi.org/10.1103/PhysRevLett.86.192}{CrossRef}]
\bibitem[Silenko and Teryaev(2005)]{Silenko:2004ad}
Silenko, A.J.; Teryaev, O.V.
\newblock {Semiclassical limit for Dirac particles interaction with a
gravitational field}.
\newblock {\em Phys. Rev. D} {\bf 2005}, {\em 71},~064016. [\href{http://dx.doi.org/10.1103/PhysRevD.71.064016}{CrossRef}]
\bibitem[Onoda et~al.(2004)Onoda, Murakami, and Nagaosa]{onoda2004hall}
Onoda, M.; Murakami, S.; Nagaosa, N.
\newblock Hall effect of light.
\newblock {\em Phys. Rev. Lett.} {\bf 2004}, {\em 93},~083901. [\href{http://dx.doi.org/10.1103/PhysRevLett.93.083901}{CrossRef}]
\bibitem[Hosten and Kwiat(2008)]{hosten2008obs}
Hosten, O.; Kwiat, P.
\newblock Observation of the spin Hall effect of light via weak measurements.
\newblock {\em Science} {\bf 2008}, {\em 319},~787--790. [\href{http://dx.doi.org/10.1126/science.1152697}{CrossRef}] [\href{http://www.ncbi.nlm.nih.gov/pubmed/18187623}{PubMed}]
\bibitem[Frolov(2024)]{Frolov:2024ebe}
Frolov, V.P.
\newblock Spinoptics in a curved spacetime.
\newblock {\em Phys. Rev. D} {\bf 2024}, {\em 110},~064020. [\href{http://dx.doi.org/10.1103/PhysRevD.110.064020}{CrossRef}]
\bibitem[Frolov and Shoom(2024)]{Frolov:2024qow}
Frolov, V.P.; Shoom, A.A.
\newblock {Gravitational spinoptics in a curved space-time}.
\newblock {\em J. Cosmol. Astropart. Phys.} {\bf 2024}, {\em 10},~039. [\href{http://dx.doi.org/10.1088/1475-7516/2024/10/039}{CrossRef}]
\bibitem[Andersson and Oancea(2023)]{Andersson:2023bvw}
Andersson, L.; Oancea, M.A.
\newblock {Spin Hall effects in the sky}.
\newblock {\em Class. Quant. Grav.} {\bf 2023}, {\em 40},~154002. [\href{http://dx.doi.org/10.1088/1361-6382/ace021}{CrossRef}]
\bibitem[Deriglazov(2021)]{Deriglazov:2021gwa}
Deriglazov, A.A.
\newblock {Massless polarized particle and Faraday rotation of light in the
Schwarzschild spacetime}.
\newblock {\em Phys. Rev. D} {\bf 2021}, {\em 104},~025006. [\href{http://dx.doi.org/10.1103/PhysRevD.104.025006}{CrossRef}]
\bibitem[Xiao et~al.(2010)Xiao, Chang, and Niu]{Xiao:2009rm}
Xiao, D.; Chang, M.C.; Niu, Q.
\newblock {Berry Phase Effects on Electronic Properties}.
\newblock {\em Rev. Mod. Phys.} {\bf 2010}, {\em 82},~1959--2007. [\href{http://dx.doi.org/10.1103/RevModPhys.82.1959}{CrossRef}]
\bibitem[Noh et~al.(2021)Noh, Alsing, Ahn, Miller, and Park]{noh2021quantum}
Noh, H.; Alsing, P.M.; Ahn, D.; Miller, W.A.; Park, N.
\newblock Quantum mechanical rotation of a photon polarization by Earth's
gravitational field.
\newblock {\em NPJ Quantum Inf.} {\bf 2021}, {\em 7},~163. [\href{http://dx.doi.org/10.1038/s41534-021-00471-6}{CrossRef}]
\bibitem[Wang(2024)]{Wang:2023bmd}
Wang, Z.L.
\newblock {Gravitational spin Hall effect of Dirac particle and the weak
equivalence principle}.
\newblock {\em Phys. Rev. D} {\bf 2024}, {\em 109},~044060. [\href{http://dx.doi.org/10.1103/PhysRevD.109.044060}{CrossRef}]
\bibitem[Czarnecki and Gao()]{comment:2025}
Czarnecki, A.; Gao, T.
\newblock Comment on ``Gravitational Spin Hall Effect of Dirac Particle and the
Weak Equivalence Principle''. \textit{Phys. Rev. D} \textbf{2025},
\textit{submitted}.
\bibitem[Foldy and Wouthuysen(1950)]{Foldy:1949wa}
Foldy, L.L.; Wouthuysen, S.A.
\newblock On the Dirac Theory of Spin 1/2 Particles and Its Non-Relativistic
Limit.
\newblock {\em Phys. Rev.} {\bf 1950}, {\em 78},~29--36. [\href{http://dx.doi.org/10.1103/PhysRev.78.29}{CrossRef}]
\bibitem[Foldy(2006)]{FoldyOriginFW}
Foldy, L.L.
\newblock Origins of the {FW} transformation: A memoir. In {\em Physics at a
Research University: Case Western Reserve 1830--1990}; Fickinger, W.J., Ed.;
Case Western Reserve University: Cleveland, \hl{OH, USA,} 
2006; pp. 347--351.
\bibitem[Liberman and Zel'dovich(1992)]{Liberman:1992zz}
Liberman, V.S.; Zel'dovich, B.Y.
\newblock {Spin-orbit interaction of a photon in an inhomogeneous medium}.
\newblock {\em Phys. Rev. A} {\bf 1992}, {\em 46},~5199--5207. [\href{http://dx.doi.org/10.1103/PhysRevA.46.5199}{CrossRef}] [\href{http://www.ncbi.nlm.nih.gov/pubmed/9908741}{PubMed}]
\bibitem[Yamamoto(2018)]{yamamoto2018spin}
Yamamoto, N.
\newblock Spin Hall effect of gravitational waves.
\newblock {\em Phys. Rev. D} {\bf 2018}, {\em 98},~061701. [\href{http://dx.doi.org/10.1103/PhysRevD.98.061701}{CrossRef}]
\bibitem[Bliokh et~al.(2015)Bliokh, Smirnova, and Nori]{bliokh2015quantum}
Bliokh, K.Y.; Smirnova, D.; Nori, F.
\newblock Quantum spin Hall effect of light.
\newblock {\em Science} {\bf 2015}, {\em 348},~1448--1451. [\href{http://dx.doi.org/10.1126/science.aaa9519}{CrossRef}]
\bibitem[Shockley and James(1967)]{Shockley67}
Shockley, W.; James, R.P.
\newblock ``Try Simplest Cases'' Discovery of ``Hidden Momentum'' Forces on
``Magnetic Currents''.
\newblock {\em Phys. Rev. Lett.} {\bf 1967}, {\em 18},~876--879. [\href{http://dx.doi.org/10.1103/PhysRevLett.18.876}{CrossRef}]
\bibitem[Griffiths(2012)]{griffiths2012resource}
Griffiths, D.J.
\newblock Resource letter EM-1: Electromagnetic momentum.
\newblock {\em Am. J. Phys.} {\bf 2012}, {\em 80},~7--18. [\href{http://dx.doi.org/10.1119/1.3641979}{CrossRef}]
\bibitem[Babson et~al.(2009)Babson, Reynolds, Bjorkquist, and
Griffiths]{babson2009hidden}
Babson, D.; Reynolds, S.P.; Bjorkquist, R.; Griffiths, D.J.
\newblock Hidden momentum, field momentum, and electromagnetic impulse.
\newblock {\em Am. J. Phys.} {\bf 2009}, {\em 77},~826--833. [\href{http://dx.doi.org/10.1119/1.3152712}{CrossRef}]
\bibitem[Eriksen and Kolsrud(1960)]{eriksen1960canonical}
Eriksen, E.; Kolsrud, M.
\newblock Canonical transformations of Dirac's equation to even forms.
Expansion in terms of the external fields.
\newblock {\em Nuovo Cim.} {\bf 1960}, {\em 18},~1--39. [\href{http://dx.doi.org/10.1007/BF02782145}{CrossRef}]
\bibitem[Nikitin(1998)]{nikitin1998exact}
Nikitin, A.G.
\newblock On exact Foldy-Wouthuysen transformation.
\newblock {\em J. Phys. A Math. Nucl. Gen.} {\bf
1998}, {\em 31},~3297--3300. [\href{http://dx.doi.org/10.1088/0305-4470/31/14/015}{CrossRef}]
\bibitem[Ram{\'i}rez and Deriglazov(2017)]{Ramirez:2017pmp}
Ram{\'i}rez, W.G.; Deriglazov, A.A.
\newblock {Relativistic effects due to gravimagnetic moment of a rotating
body}.
\newblock {\em Phys. Rev. D} {\bf 2017}, {\em 96},~124013. [\href{http://dx.doi.org/10.1103/PhysRevD.96.124013}{CrossRef}]
\bibitem[Newton and Wigner(1949)]{Newton:1949cq}
Newton, T.D.; Wigner, E.P.
\newblock Localized States for Elementary Systems.
\newblock {\em Rev. Mod. Phys.} {\bf 1949}, {\em 21},~400--406. [\href{http://dx.doi.org/10.1103/RevModPhys.21.400}{CrossRef}]
\bibitem[Fradkin and {Good Jr.}(1961)]{fradkin1961electron}
Fradkin, D.M.; Good, R.H., Jr.
\newblock Electron polarization operators.
\newblock {\em Rev. Mod. Phys.} {\bf 1961}, {\em 33},~343--352. [\href{http://dx.doi.org/10.1103/RevModPhys.33.343}{CrossRef}]
\bibitem[Deriglazov and Guzm{\'a}n~Ram{\'i}rez(2017)]{Deriglazov:2017jub}
Deriglazov, A.A.; Guzm{\'a}n~Ram{\'i}rez, W.
\newblock {Recent progress on the description of relativistic spin: Vector
model of spinning particle and rotating body with gravimagnetic moment in
General Relativity}.
\newblock {\em Adv. Math. Phys.} {\bf 2017}, {\em 2017},~7397159. [\href{http://dx.doi.org/10.1155/2017/7397159}{CrossRef}]
\end{thebibliography}
\end{document}